\input{aipcheck}

\documentclass[
    ,final            
  ]
  {aipproc}

\layoutstyle{6x9}

\newcommand{\lsim}{\raisebox{-0.13cm}{~\shortstack{$<$ \\[-0.07cm] $\sim$}}~}
\newcommand{\gsim}{\raisebox{-0.13cm}{~\shortstack{$>$ \\[-0.07cm] $\sim$}}~}
\begin{document}

\title{The dark matter as a light gravitino (II)${}^1$ } \footnote{based on
work 
in collaboration with
K. Jedamzik ({\sl LPTA-Montpellier}), M. Lemoine ({\sl IAP-Paris}) \cite{JLM04,
JLM05}; and  work in progress, M. Kuroda (Meiji-Gakuin), M. Lemoine (Paris), M. Capdequi-Peyranère
(Montpellier).}


\author{Gilbert \sc Moultaka}{
  address={Laboratoire de Physique Th\'eorique et Astroparticules \\
{\sl UMR5207--CNRS}, Universit\'e Montpellier II \\
Place E. Bataillon, F--34095 Montpellier Cedex 5, France}
}



\begin{abstract}
We address the question of gravitino dark matter in the context of gauge 
mediated supersymmetry breaking models. 
\end{abstract}

\maketitle


\section{Introduction}

In scenarios where supersymmetry breaking is triggered by non-perturbative dynamics of 
some (secluded) gauge sector and communicated to the MSSM by a messenger sector through 
perturbative gauge interactions,  
the susy breaking scale $\sqrt{F}$ and the mass scale $\Lambda$ of the 
secluded gauge sector can be well below the Planck scale. Recent developments 
\cite{ISS} stressing the existence of metastable susy breaking vacua,  
have renewed the interest in such gauge-mediated susy breaking (GMSB) scenarios
opening new possibilities for the model-building  \cite{GR99}, and appear to be 
very interesting from the early Universe point of view as well \cite{AK}. 
On the other hand, the gravitational interactions which play 
a minor role for susy breaking in GMSB models remain  
physically relevant through the coupling to supergravity, at least in order 
to absorb  the unphysical goldstino component, to adjust the cosmological 
constant to a small value and to avoid a massless R-axion. 
Moreover, if the above mentioned two scales combine to trigger the electroweak 
symmetry breaking yielding $G_F^{-1/2} \sim (\alpha/4 \pi) k {F / \Lambda } $, where 
$G_F$ is Fermi's constant (and $0< k \le 1$ measures the secludedness of 
the secluded  sector), then the gravitino mass 
$m_{3/2} \simeq F/(\sqrt{3} m_{\rm Pl}) \sim \left(4 \pi/ \alpha)(\Lambda /\sqrt{3} k m_{\rm Pl}\right) 
G_F^{-1/2}$ where $m_{\rm Pl}$  is the reduced Planck mass, is expected to be 
very small ($\lsim {\cal O}(1)$ GeV) and is the lightest supersymmetric particle
(LSP). The question then arises as to which particle can be a good candidate 
 for the cold dark matter (CDM) in this case? To answer this question requires an
unconventional treatment as compared to the Neutralino ``vanilla'' candidate
or even to the heavy gravitino candidate in the context of gravity mediated susy
breaking models. Indeed, in contrast with the latter where the hidden sector 
is typically too heavy to be produced at the end of inflation, the secluded and 
messenger sectors of GMSB provide stable particles that may be present in the early Universe for a
sufficiently heavy reheat temperature $T_{RH}$. We consider hereafter such
configurations assuming that only the messenger (including the spurion) sector 
can be produced and  illustrate its relevance to the issue of the CDM.

\section{A Messenger Solution to the Gravitino Problem}

The mass degeneracy within a supermultiplet of messenger fields is lifted by 
susy breaking leading to a lighter and a heavier scalar messengers with masses
$M_\pm =M_X(1 \pm {k F/M_X^2})^{1/2}$ and a fermionic partner with mass 
$M_X$ (where $F$ and $M_X$ are related to the dynamical scale $\Lambda$).
 Thus ${k F/M_X^2} < 1$. Moreover, one has to require 
 ${k F/M_X} \lsim 10^5 \mbox{GeV}$ to ensure an MSSM spectrum 
 $\lsim {\cal O}(1) \mbox{TeV}$. One then expects typically 
 $M_X \gsim 10^5 \mbox{GeV}$. In typical GMSB models the lightest messenger 
 particle (LMP) with mass $M_{-}$ is stable due to the conservation of a 
 messenger quantum number. If present in the early Universe the messenger 
 particles are thermalized through their gauge interactions with the thermal 
bath. The corresponding LMP relic density is calculable similarly to that of 
the Neutralino LSP.  However, it turns out to be typically too large to account 
for the CDM (albeit fine-tuning) even in the most favorable case of the 
electrically neutral component of a $\mathbf{5}+\mathbf{\overline{5}}$ 
representation of $SU(5)$ where it is found to scale as 
$\Omega_M h^2  \simeq 10^5 \left({M_{-}/10^3 TeV}\right)^2$ 
with the LMP mass \cite{DGP}. The situation is even worse in the case of
$SO(10)$ where the LMP is an MSSM singlet with a suppressed annihilation 
cross-section leading to a very large relic density. One possible cure to this 
messenger overcloser problem, namely to allow the LMP to decay, can actually 
turn out to be a blessing regarding a solution to the gravitino problem and 
simultaneously letting the gravitino account for the CDM in the context of GMSB
 models.\footnote{One can easily argue for
an unstable LMP once the GMSB model is coupled to 
supergravity, invoking the violation of the messenger number conservation 
by gravitational interactions akin to discrete accidental symmetries.   
The resulting messenger number violating operators are then Planck scale 
suppressed and would not upset the natural suppression of the flavor 
changing neutral currents in GMSB models.}
The LMP late decay into MSSM particles can release enough entropy
to dilute the initial gravitino relic density down to a level which can 
account for the CDM in the Universe even for very high $T_{RH}$,
\cite{FY02, JLM04, JLM05}.
For this to work, though, the LMP should dominate the Universe 
energy density before it decays, and should decay after the gravitino has 
freezed-out from the thermal bath. The necessary condition $T_d < T_{MD} <
T^f_{3/2}$ [where $T_d, T_{MD}, T^f_{3/2}$ denote respectively the LMP decay
and messenger matter domination temperatures, and the gravitino freeze-out temperature] 
is then determined by the particle properties and annihilation cross-section 
and decay width of the LMP, delineating the favorable parts of the parameter 
space. We have studied this scenario in detail for the case of $SU(5)$ \cite{JLM04}
and $SO(10)$ \cite{JLM05}. In the next section we concentrate on the latter case
with one set of messengers transforming as  
$\mathbf{16} + \overline{\mathbf{16}}$.

\begin{figure}[t]
\label{fig1}
  \vspace{-5cm}
  \includegraphics[width=.8 \textwidth, height=.8 \textheight, keepaspectratio,
bb= -15 460 580 660, clip]{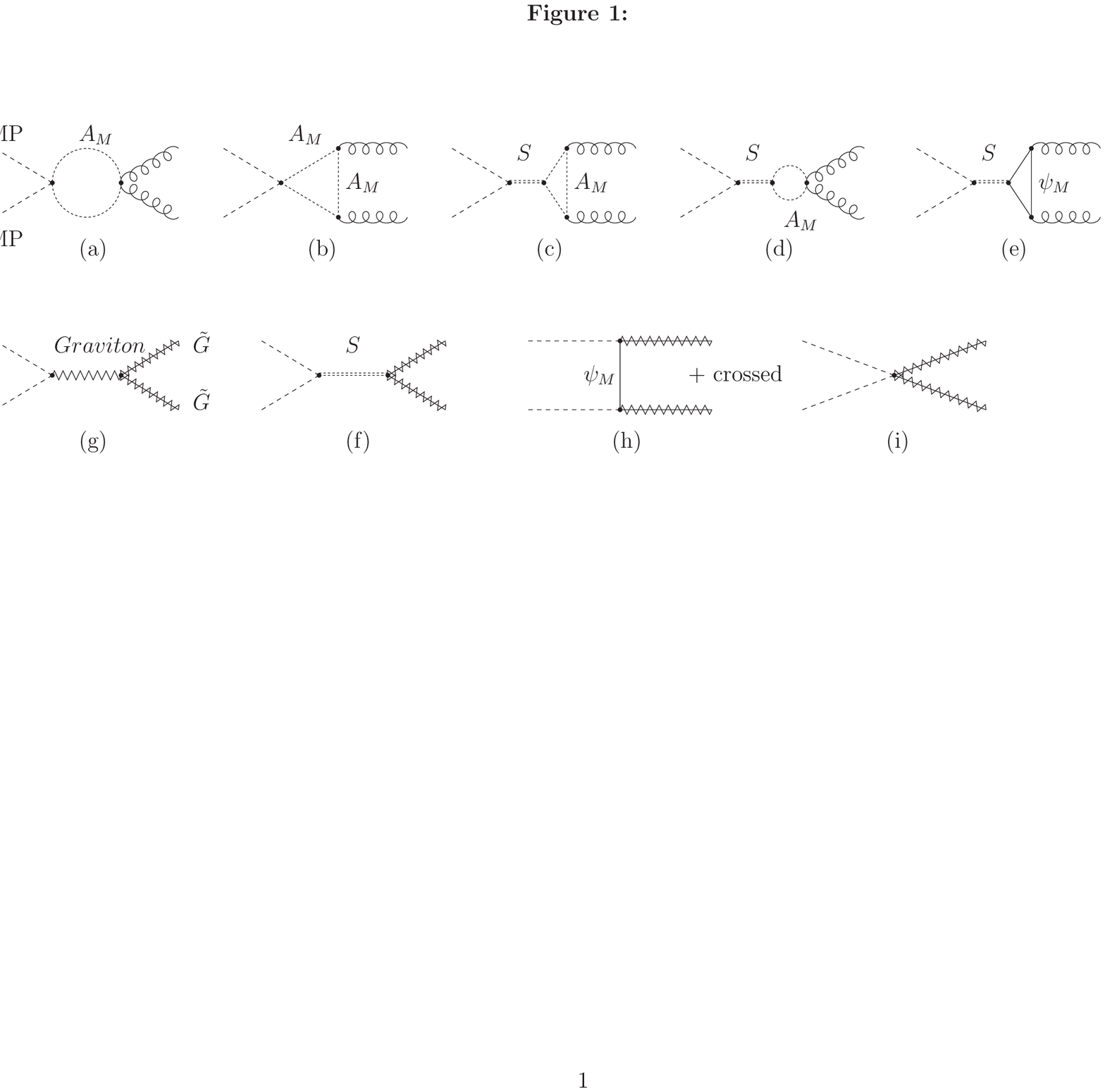}
  \caption{Feynman diagrams of the leading LMP annihilation into $2$
  gluons (a-e) or $2$ gravitinos (f-i).}
\end{figure}
 
\section{Gravitino relic abundance}

\begin{figure}[t]
\label{fig2}
  \includegraphics[height=.3\textheight]{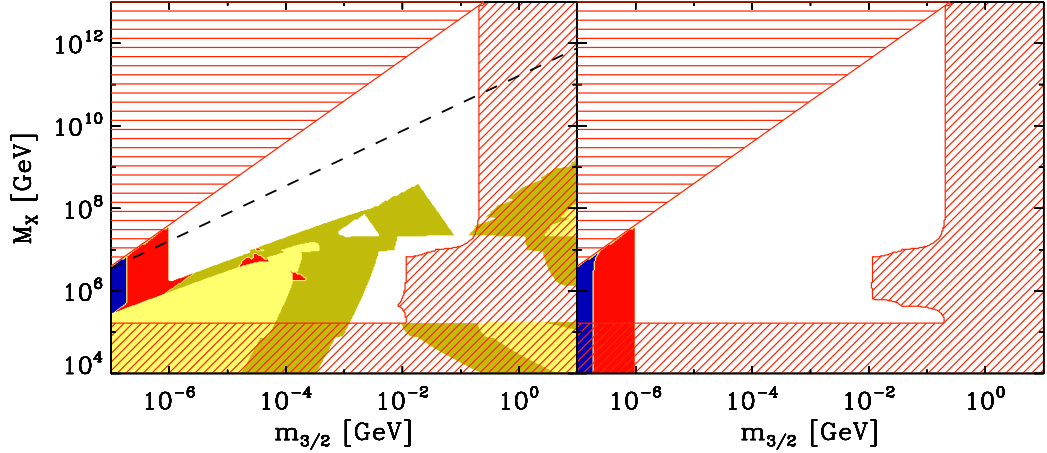}
  \caption{Contours of $\Omega_{3/2}$ in the plane $M_X (\equiv M_-)-m_{3/2}$
  for one pair of messengers sitting in
  $\mathbf{16}+\overline{\mathbf{16}}$ representations of $SO(10)$;
  the LMP is a singlet under $SU(3)\times
  SU(2)\times U(1)$. We take for illustration
  $\kappa^2 \simeq \alpha_s/4\pi$, $f\sim {\cal O}(1)$ and $f' \simeq 5.
  10^{-2}$ and a bino NLSP with $M_{\rm NLSP}=150\,$GeV, $M_{gluino}=1\,$TeV 
  and $k F/M_- \simeq 10^5$GeV;  
  blue (hot), red (warm), green (cold) DM with $0.01 < \Omega_{grav}
  <1$; yellow ($\Omega_{grav} < 0.01$), white ($\Omega_{grav} > 1$). In the 
  right (left) panel the spurion is lighter (heavier) 
  than the messenger. (taken from \cite{JLM04}.)}
\end{figure}

The entropy release $\Delta S \equiv S_{\rm after}/S_{\rm before}$, diluting the initial
gravitino density, is determined by the temperatures before and after LMP decay
and can be approximated to $T_{MD}/T_d$.  $T_{MD}$ is given by the LMP yield 
and mass ($T_{MD}\simeq (4/3)M_{-} \times Y_{\rm LMP}$) and $T_d$ is determined
by the LMP width ($\Gamma_{\rm LMP} \simeq H(T_d)$). $Y_{\rm LMP}$ is
determined by the LMP annihilation into MSSM particles. Since in our case 
the LMP is an $SU(5)$ singlet \cite{DGP, JLM05}, this annihilation proceeds
via loop effects of virtual messengers ($A_M, \psi_M$) 
and spurion ($S$) exchange,  fig.\ref{fig1}. We 
parameterize the thermally averaged leading annihilation cross-section into $2$ gluons 
as $\langle \sigma_{1{\rm loop}}v\rangle \sim f (\alpha_s/4\pi)^2 \kappa^4/s$ 
where $\kappa$ is the spurion-messenger coupling
($W \supset   \kappa \hat{S} \mathbf{16}_M\overline{\mathbf{16}}_M$),
$\alpha_s$ the strong coupling constant,
$\sqrt{s}$ the C.M. energy and $f$ a form factor depending on the internal
masses.  
The LMP decay is induced by Planck scale suppressed
non-renormalizable messenger number violating operators which can originate 
from the K\"ahler potential, e.g. $K \supset \mathbf{16}_F 
\overline{\mathbf{16}}_M^\dag \mathbf{10}_H/m_{\rm Pl}$, 
 or from the superpotential, e.g. $W \supset 
\overline{\mathbf{16}}_M {\mathbf{16}}_F {\mathbf{16}}_F \mathbf{10}_H /m_{\rm Pl}$,
leading respectively to $2$- and $3$-body decays,
where $ \mathbf{16}_M (\overline{\mathbf{16}}_M), \mathbf{16}_F$ and 
$\mathbf{10}_H$ denote respectively the messenger, the standard matter
and the  electroweak Higgs supermultiplets.
We assume a typical decay width
$\Gamma_{\rm LMP} = (1/16\pi) f' M_X^3/m_{\rm Pl}^2$ where $f'$ 
parameterizes our ignorance of the couplings  and possible further phase space 
suppression. When the necessary temperature conditions are met,  the final 
gravitino relic density is given by $\Omega_{{}_{grav}} = \Omega_{{}_{grav}}^{th}/\Delta S + 
\Omega_{{}_{grav}}^{{}^{Mess}} + \Omega_{{}_{grav}}^{{}^{NLSP}}$ where the last
two contributions denote non-thermal production through late decays or
scattering. One should also consider various cosmological constraints 
(hotness/warmness, BBN, species dilution, etc...). In fig.\ref{fig2} we
illustrate the case with $T_{RH} \simeq 10^{12}$ GeV, see also \cite{JLM04}. 
The horizontal red shading shows the theoretically excluded region where 
$ k > 1$; the other red shading indicates the region excluded by BBN 
constraints. If the spurion is heavier than the LMP, gravitino cold DM 
(green region) occurs for relatively light
LMPs and $m_{3/2} \sim 1 \mbox{keV} - 10 \mbox{MeV}$. More generally, in 
the models of ref. \cite{GMSB} one finds \cite{JLM05}
$\Omega_{grav}h^2 \,\simeq\, 10^3
f^{0.8} \kappa^{3.2} f'^{1/2}\left({M_{-}/ 10^6\,{\rm
GeV}}\right)^{-0.3}\left({m_{3/2}/1\,{\rm MeV}}\right)$ for non-relativistic
LMP freeze-out, putting the gravitino relic abundance in the ballpark of WMAP 
results, for $\kappa \sim {\cal
O}(10^{-1})$ and typical ranges for $f$ and $f'$. The LMP can also annihilate
into 2 gravitinos through gravitational interactions, fig.\ref{fig1}. For
very heavy spurions the annihilation cross-section at rest reads 
$\langle \sigma v\rangle\,\simeq\, (1/ 24\pi) k^2 M_-^2/(m_{3/2} m_{\rm Pl})^2
$. It can dominate the 1-loop annihilation, eventually saturating the unitarity
limit (the black dashed line in fig.\ref{fig2}), thus disfavouring gravitino 
CDM solutions for a very heavy LMP.

To summarize, a light gravitino can account for  CDM  irrespective of
$T_{RH}$, making it a good DM candidate in GMSB: typically  if $T_{RH} \lsim 10^5 \mbox{GeV}$ then the messengers are 
not produced and thermal gravitinos with $m_{3/2} \simeq 1 \mbox{MeV}$ provide
the right CDM density, while for $T_{RH} \gsim 10^5 \mbox{GeV}$ the messenger
can be present and should be unstable, thus providing a source of entropy 
production that can reduce a thermally overproduced gravitino to a
cosmologically acceptable level.
Moreover, various constraints (e.g. on $T_{RH}$, \cite{SP}, or on 
the gravitino mass \cite{VLHMR}) simply do not apply in the scenarios we 
have illustrated, thus escaping  possible tension with thermal leptogenesis.
 



\end{document}